\UseRawInputEncoding
\documentclass[aps,prb,twocolumn,showpacs,psfig,superscriptaddress,longbibliography]{revtex4-2}
\usepackage{amsfonts}
\usepackage{mathrsfs}
\usepackage{amsmath}
\usepackage{color}
\usepackage{natbib}
\usepackage{textcomp}
\usepackage{graphicx}
\usepackage{bm}
\usepackage{amssymb}

\usepackage{xspace}
\usepackage{epstopdf}
\usepackage{dcolumn}
\usepackage{longtable}
\usepackage{multirow}
\usepackage[colorlinks=true, letterpaper=true, pdfstartview=FitV, linkcolor=blue, citecolor=blue, urlcolor=blue]{hyperref}
\usepackage{float}

\makeatletter

\newcommand{\Rmnum}[1]{\expandafter\@slowromancap\romannumeral #1@}
\makeatother

\begin{document}

\title{Enhanced superconductivity in C-S-H compounds at high pressure}

 \author{Zheng-Wei Liao}
 \affiliation{School of Physical Sciences, University of Chinese Academy of Sciences, Beijing 100049, China}

 \author{Zhen Zhang}
 \affiliation{School of Physical Sciences, University of Chinese Academy of Sciences, Beijing 100049, China}

 \author{Jing-Yang You}
\email{phyjyy@nus.edu.sg}
\affiliation{Department of Physics, National University of Singapore, 2 Science Drive 3, Singapore 117551}

 \author{Bo Gu}
 \email{gubo@ucas.ac.cn}
 \affiliation{Kavli Institute for Theoretical Sciences, and CAS Center for Excellence in Topological Quantum Computation, University of Chinese Academy of Sciences, Beijng 100190, China}

 \author{Gang Su}
 \email{gsu@ucas.ac.cn}
 \affiliation{School of Physical Sciences, University of Chinese Academy of Sciences, Beijing 100049, China}
 \affiliation{Kavli Institute for Theoretical Sciences, and CAS Center for Excellence in Topological Quantum Computation, University of Chinese Academy of Sciences, Beijng 100190, China}

\begin{abstract}
Recently, the superconducting transition temperature Tc = 287 K has been experimentally obtained in the material composed of carbon, sulfur, and hydrogen under the high pressure of 267 GPa.  The material structure is unknown yet, where the carbon and sulfur were added at a molar ratio of 1:1. Here, fixing the molar ratio of C : S = 1:1, we studied several possible C-S-H structures, and found a new stable structure C$_{2}$S$_{2}$H$_{4}$ using the first-principles calculations. The C$_2$S$_2$H$_4$ shows an insulator-to-metal transition and the superconducting ground state at the pressure of 64 GPa, and its Tc can reach 16.5 K at 300 GPa. In addition, we found another stable structure of C$_2$S$_3$H$_4$, whose Tc is 47.4 K at 300 GPa. The calculations show that the added S atom in C$_{2}$S$_{3}$H$_{4}$ breaks part of C-H bonds in C$_{2}$S$_{2}$H$_{4}$, makes the vibration of H atom at a lower frequency, and thus enhances the electro-phonon coupling and Tc. Our results suggest that the molar ratio of C:S lower than 1:1 in the C-S-H systems may be favorable to enhance Tc. This can be useful to figure out the structure of the C-S-H material with room temperature Tc in the recent experiment. 

\end{abstract}
\pacs{}
\maketitle


Room temperature superconductors have been the long-time dream of scientists because of the potential applications in the fields of information, detection, transportation, and power technology. Since the discovery of superconductivity in 1911~\cite{KamerlinghOnnes1911}, great achievements have been made in theory and experiment. In 1957, Bardeen, Cooper and Schrieffer have proposed the BCS theory~\cite{Bardeen1957}, which can well interpret many traditional superconductors. Until 1986, the highest reported transition temperature Tc of superconducting materials was only 23.2 K, which greatly limited the practical application of superconducting materials. In 1986, the Tc of 35 K was observed in a copper oxide LaBaCuO~\cite{Bednorz1986}, which broke the record of Tc and attracted great world wide attention. The discovery of superconductor YBa$_2$Cu$_3$O$_{7-y}$ with a Tc more than 90 K makes the Tc beyond the liquid nitrogen temperature region (77 K) for the first time~\cite{Wu1987}. In 1988, the superconductor Bi$_2$Sr$_2$Ca$_2$Cu$_3$O$_{10+\delta}$ with a Tc of 105 K~\cite{Maeda1988} and the superconductor Tl$_2$Ba$_2$Ca$_2$Cu$_3$O$_{10}$ with a Tc of 125 K were discovered~\cite{Sheng1988}. In 1993, a high Tc of 130 K was obtained in the superconductor HgBa$_2$Ca$_2$Cu$_3$O$_{8+\delta}$~\cite{Schilling1993}. In 2008, the layered material LaFeAsO with a Tc of 26 K~\cite{Kamihara2008} opened the exploration of iron-based superconductors. Then a series of iron-based superconductors, such as SmFeAsO$_{0.85}$F$_{0.15}$ with a Tc of 43 K~\cite{Chen2008}, CeFeAsO$_{1-x}$F$_x$ with a Tc of 41 K~\cite{Chen2008a}, NdFeAsO$_{0.85}$F$_{0.15}$ with a Tc of 51 K, and PrFeAsO$_{0.85}$F$_{0.15}$ with a Tc of 50 K~\cite{Ren2008} were observed. In addition, the alkali metal doped fullerenes~\cite{Hebard1991,Haddon1991,Rosseinsky1991}, the heavy fermion system containing radioactive elements~\cite{Steglich1979}, MgB$_2$~\cite{Nagamatsu2001}, Na-doped T-carbon~\cite{You2020} and other superconducting materials have been reported theoretically or experimentally~\cite{Tanaka2017,Cui2020,Sun2020,Cataldo2021,You2021}.

Following the BCS theory, it is highly expected that the metallic hydrogen may have high Debye temperature, strong electron-phonon coupling (EPC), and thus high Tc~\cite{Ashcroft1968,Bonev2004,McMahon2011,McMinis2015}. In 1935, Wigner and Huntington proposed theoretically that hydrogen could be transformed into metal state under extreme conditions of high pressure~\cite{Wigner1935}.  
In 2004, Ashcroft pointed out that high-temperature superconductors can be obtained by pressurizing hydrogen rich materials~\cite{Ashcroft2004}. In 2014, H$_3$S was predicted to become a hydride superconductor under high pressure~\cite{Li2014,Duan2014}, and soon after a Tc of 203 K at 90 GPa was obtained experimentally~\cite{Drozdov2015}. Higher Tc was theoretically proposed in the P-, C- and Si-doped H$_3$S~\cite{Ge2016,Ge2020}. In recent years, many new hydrogen-rich materials with different structures were predicted, such as lanthanum, yttrium hydrides~\cite{Peng2017,Liu2017,KwangHua2020} and scandium hydrides~\cite{Ye2018}. The lanthanum hydride LaH$_{10}$ was synthesized and demonstrated to exhibit a Tc of 250-260 K at 170-190 GPa~\cite{Somayazulu2019,Drozdov2019,Hong2020}. Yttrium superhydride YH$_9$ with a Tc of 262 K at 182 $\pm$ 8 GPa was also obtained~\cite{Snider2021}.
In 2020, the room temperature superconductor with Tc = 287 K at 267 GPa has been experimentally obtained in the material composed of carbon, sulfur, and hydrogen~\cite{Snider2020}. In the experiment, the carbon and sulfur were added at a molar ratio of 1:1, but the material structure is unknown yet. 

In this Letter, by fixing the molar ratio of C : S = 1:1, we studied several possible C-S-H structures, and found a stable structure C$_2$S$_2$H$_4$ by first-principle calculations. A Tc of 16.5 K at 300 GPa is predicted in C$_2$S$_2$H$_4$. In addition, a higher Tc of 47.4 K at 300 GPa is obtained in another stable material C$_2$C$_3$H$_4$. The added S atom in C$_2$S$_3$H$_4$ makes the vibration of H atoms at a lower frequency, and thus enhances the electron-phonon coupling and Tc. Our results suggest that the molar ratio of C:S lower than 1:1 in the C-S-H systems may be favorable to enhance Tc.

\begin{figure}[!htbp]
	\centering
	\includegraphics[scale=0.25,angle=0]{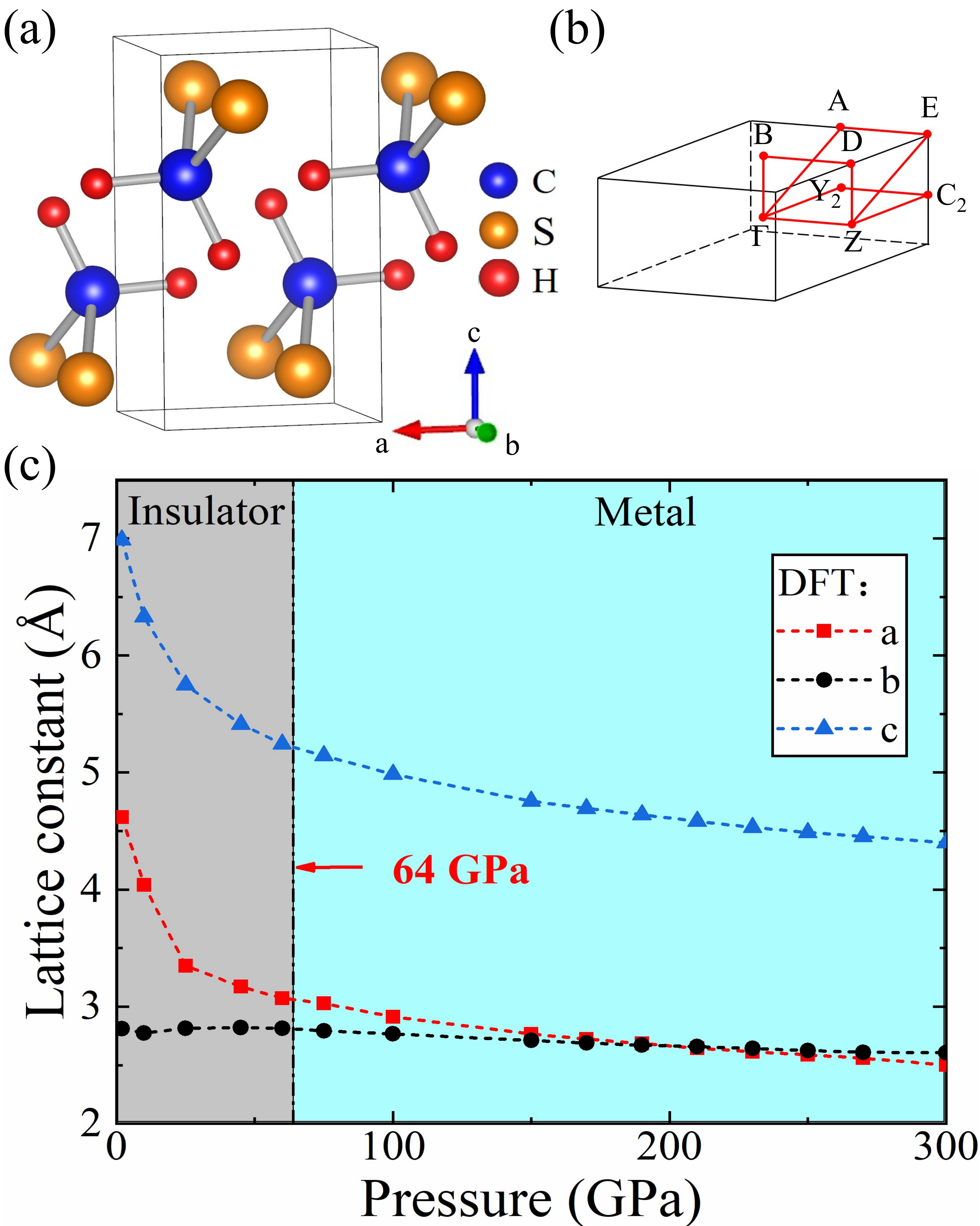}\\
	\caption{(a) The crystal structure and (b)  Brillouin zone (BZ) of  C$_{2}$S$_{2}$H$_{4}$. (c) The pressure dependence of lattice paremeters in C$_{2}$S$_{2}$H$_{4}$ obtained by the DFT calculations, where the critical pressure of insulator to metal phase transition is 64 GPa.}\label{fig1}
\end{figure}

In the experiment of the room temperature superconductor C-S-H system, carbon and sulfur were mixed in a molar ratio of 1:1, and hydrogen was introduced under a certain pressure for reaction\cite{Snider2020}. Because the material structure is unknown yet, we started to search for the possible stable structures by fixing the ratio of C : S = 1:1. Though many proposed C-S-H structures are unstable in the calculations, we obtained a stable structure C$_{2}$S$_{2}$H$_{4}$ which contains 8 atoms in the primitive cell. The C$_{2}$S$_{2}$H$_{4}$ at 300 GPa belongs to the monoclinic crystal with the space group of $P2_1/m$ (No. 11). The optimized lattice constants are $a = 2.501$ \AA, $b = 2.606$ \AA, and $c = 4.401$ \AA~ with lattice angles of $\alpha = 90^{\circ}$, $\beta = 90.608^{\circ}$, and $\gamma = 90^{\circ}$ as shown in Fig.~\ref{fig1}(a). The pressure dependent lattice constants of C$_2$S$_2$H$_4$ is plotted in Fig.~\ref{fig1}(c). It is noted that the lattice constants $a$ and $c$ increase in a similar trend with the decrease of pressure, while $b$ slightly changes. The lattice constants of C$_{2}$S$_{2}$H$_{4}$ at normal pressure become  $a = 4.619$ \AA, $b = 2.813$ \AA, and $c = 6.897$ \AA, and the lattice angles are $\alpha = 90^{\circ}$, $\beta = 103.118^{\circ}$, and $\gamma = 90^{\circ}$. The change of C$_{2}$S$_{2}$H$_{4}$ crystal structure will dramatically affect its electronic and many other properties.

\begin{figure}[!htbp]
  \centering
  \includegraphics[scale=0.27,angle=0]{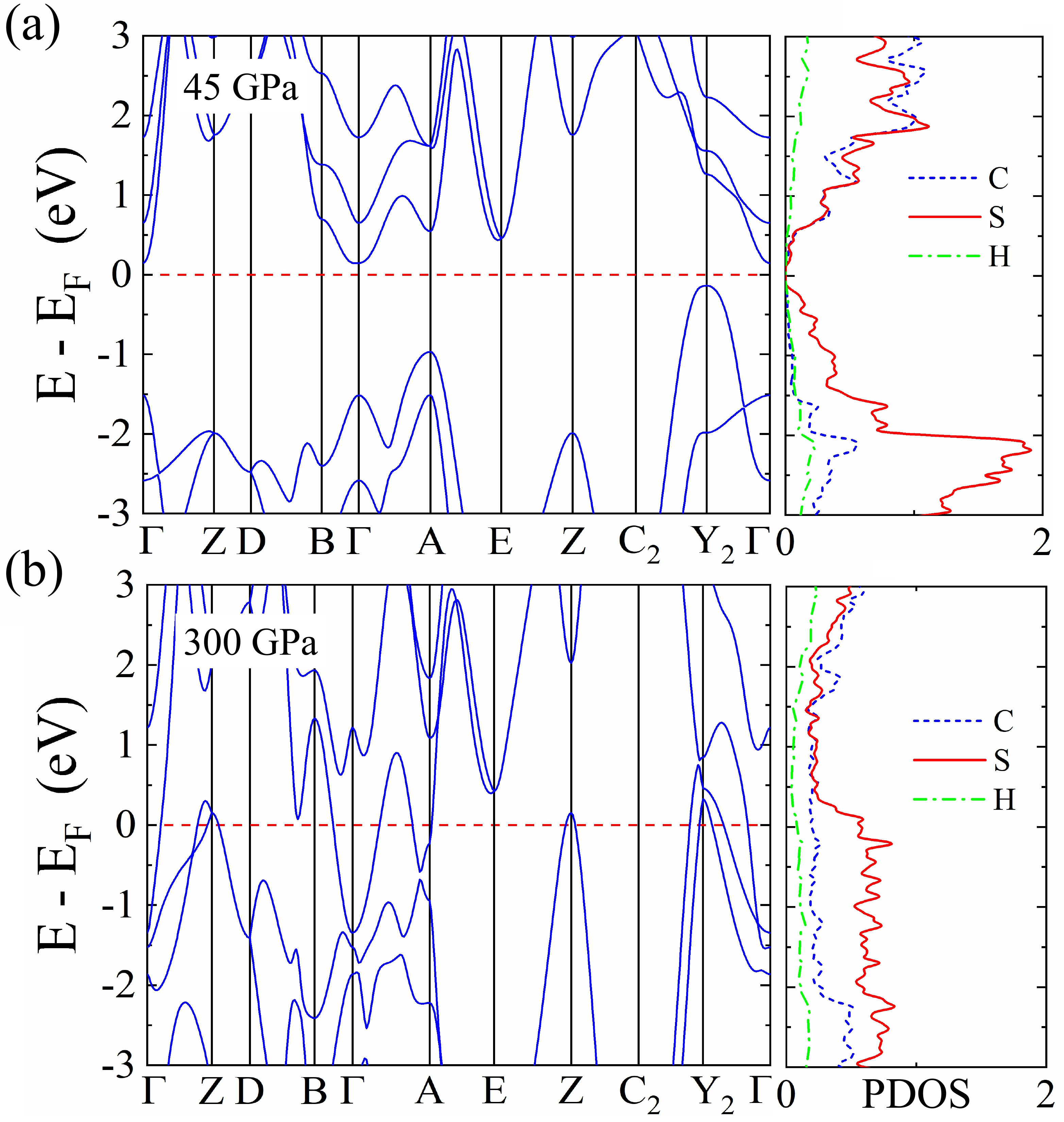}\\
  \caption{The electron band structures and projected density of states (PDOS) of C$_{2}$S$_{2}$H$_{4}$ at (a) 45 and (b) 300 GPa. The Fermi level is indicated by dash lines.}\label{fig2}
\end{figure}

\begin{figure}[!htbp]
	\centering
	\includegraphics[scale=0.27,angle=0]{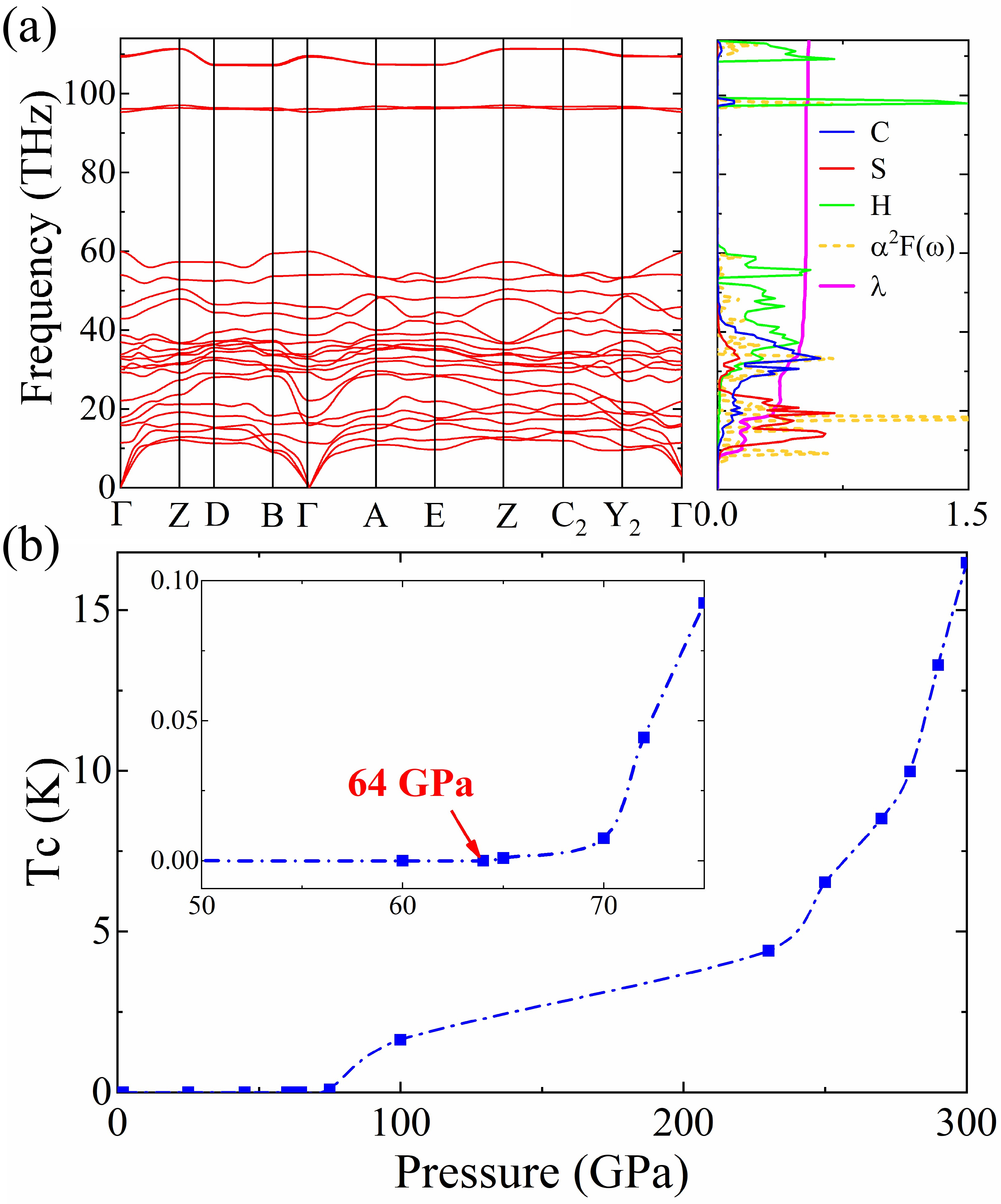}\\
	\caption{(a) The phonon spectra, projected phonon density of states, Eliashberg spectral function $\alpha^2F(\omega)$, and the cumulative frequency-dependent of EPC $\lambda(\omega)$ of C$_{2}$S$_{2}$H$_{4}$ at 300 GPa. (b) Pressure dependent Tc, where the critical pressure of insultor to metal (superconductor) phase transition is indicated.}\label{fig3}
\end{figure}

In Fig.~\ref{fig2}, we plot the electron band structures and projected density of states (PDOS) at two representative pressures of 45 and 300 GPa. We find that there is an energy gap of 0.283 eV in C$_{2}$S$_{2}$H$_{4}$ at 45 GPa, while it disappears at 300 GPa, which indicates the existence of the phase transition between insulator and metal as the change of pressure. In fact, we find that the critical pressure of insulator-metal phase transition is 64 GPa , as indicated in Fig.~\ref{fig1}(c). In addition, it is noted that the DOS near Fermi level at 300 GPa mainly comes from the contribution of S atoms. 

\begin{figure}[!htbp]
	\centering
	\includegraphics[scale=0.35,angle=0]{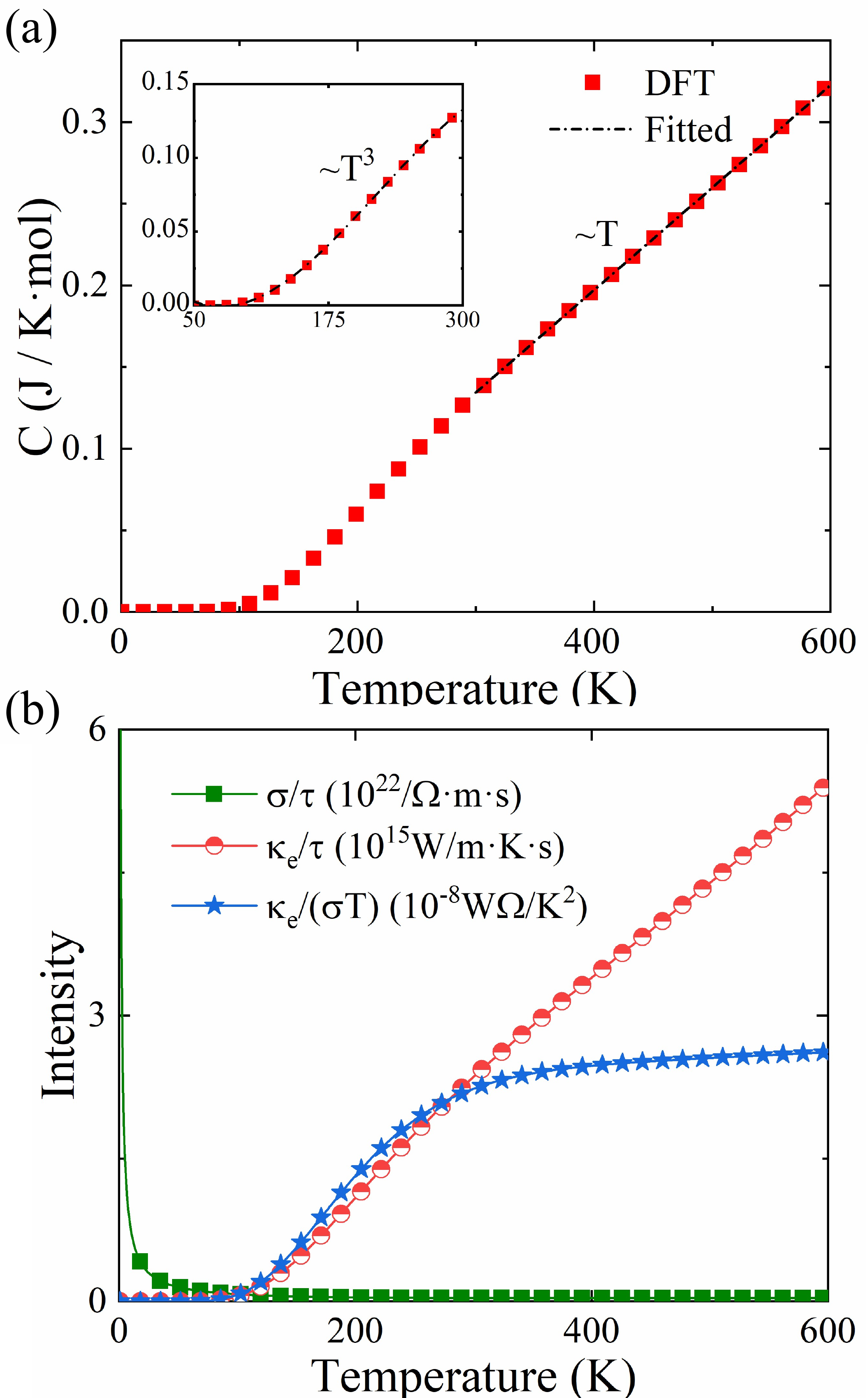}\\
	\caption{(a) Temperature dependence of electronic specific heat C of C$_{2}$S$_{2}$H$_{4}$ at 300 GPa. The upper left inset is the enlarged parts of the specific heat at temperature 50 K $<$ T $<$ 300 K. The dash lines are fitting curves at different temperature regions. (b) Temperature dependent Lorenz number $L$ [$L=\kappa_e$/($\sigma$T)], the temperature dependent electrical $\sigma$ and thermal $\kappa_e$ conductivities over the relaxation time $\tau$. }\label{fig5}
\end{figure}

The metallicity of C$_2$S$_2$H$_4$ at 300 GPa allows to investigate its possible superconductivity. Fig~\ref{fig3}(a) gives the phonon spectra along high-symmetry paths $\Gamma$-$Z$-$D$-$B$-$\Gamma$-$A$-$E$-$Z$-$C_2$-$Y_2$-$\Gamma$. There is no imaginary frequency mode in phonon spectra, indicating the dynamic stability of C$_{2}$S$_{2}$H$_{4}$. Meanwhile, the phonon density of states (PhDOS) in Fig.~\ref{fig3}(a) shows that S atoms vibrate at low frequency from 10 to 20 THz, C atoms vibrate around 30 THz, and H atoms vibrate at relative high frequency. Based on Migdal-Eliashberg theory~\cite{Guentherodt1980,Giustino2017}, the Eliashberg spectral function $\alpha^2F(\omega)$, and the cumulative frequency-dependent EPC $\lambda(\omega)$ of C$_{2}$S$_{2}$H$_{4}$ at 300 GPa are calculated as also given in Fig.~\ref{fig3}(a). It is obvious that S atoms located at low frequency make more contributions to the EPC, while H atoms at high frequency have little contribution to EPC.

Based on the BCS theory with the McMillan-Allen-Dynes approach~\cite{Allen1975,McMillan1968} and taking a typical value of the effective screened Coulomb repulsion constant $\mu^*$ = 0.1, we obtain the Tc of C$_{2}$S$_{2}$H$_{4}$ at different pressures as plotted in Fig.~\ref{fig3}(b). We find that with the increase of pressure, C$_{2}$S$_{2}$H$_{4}$ does not become superconducting until 64 GPa because of the insulator-to-metal phase transition, and the Tc can reach 16.475 K at 300 GPa.

\begin{table}
	\renewcommand\arraystretch{1.25}
	\caption{The Tc and total EPC $\lambda$ of C$_{2}$S$_{2}$H$_{4}$, C$_{2}$SClH$_{4}$, C$_{2}$SPH$_{4}$, C$_{2}$SFH$_{4}$, C$_{2}$S$_{3}$H$_{4}$ at 300 GPa.}
	\begin{tabular}{l<{\centering}p{2.4cm}<{\centering}p{2.4cm}<{\centering}p{2.4cm}<{\centering}}
		\hline
		\hline
		& C$_{2}$S$_{2}$H$_{4}$ & C$_{2}$SClH$_{4}$ & C$_{2}$SPH$_{4}$  \\
		\hline
		Tc (K)             & 16.475 & 8.532   & 0.155   \\
		$\lambda$          & 0.5504 & 0.5130  & 0.2712   \\
		\hline
		\hline
		& C$_{2}$SFH$_{4}$ & C$_{2}$S$_{3}$H$_{4}$ \\
		\hline
		Tc (K)             & 3.613  & 47.439   \\
		$\lambda$          & 0.3986 & 0.8646     \\
		\hline
		\hline
		
	\end{tabular}
	\label{T-2}
\end{table}	

\begin{figure}[!htbp]
	\centering
	\includegraphics[scale=0.28,angle=0]{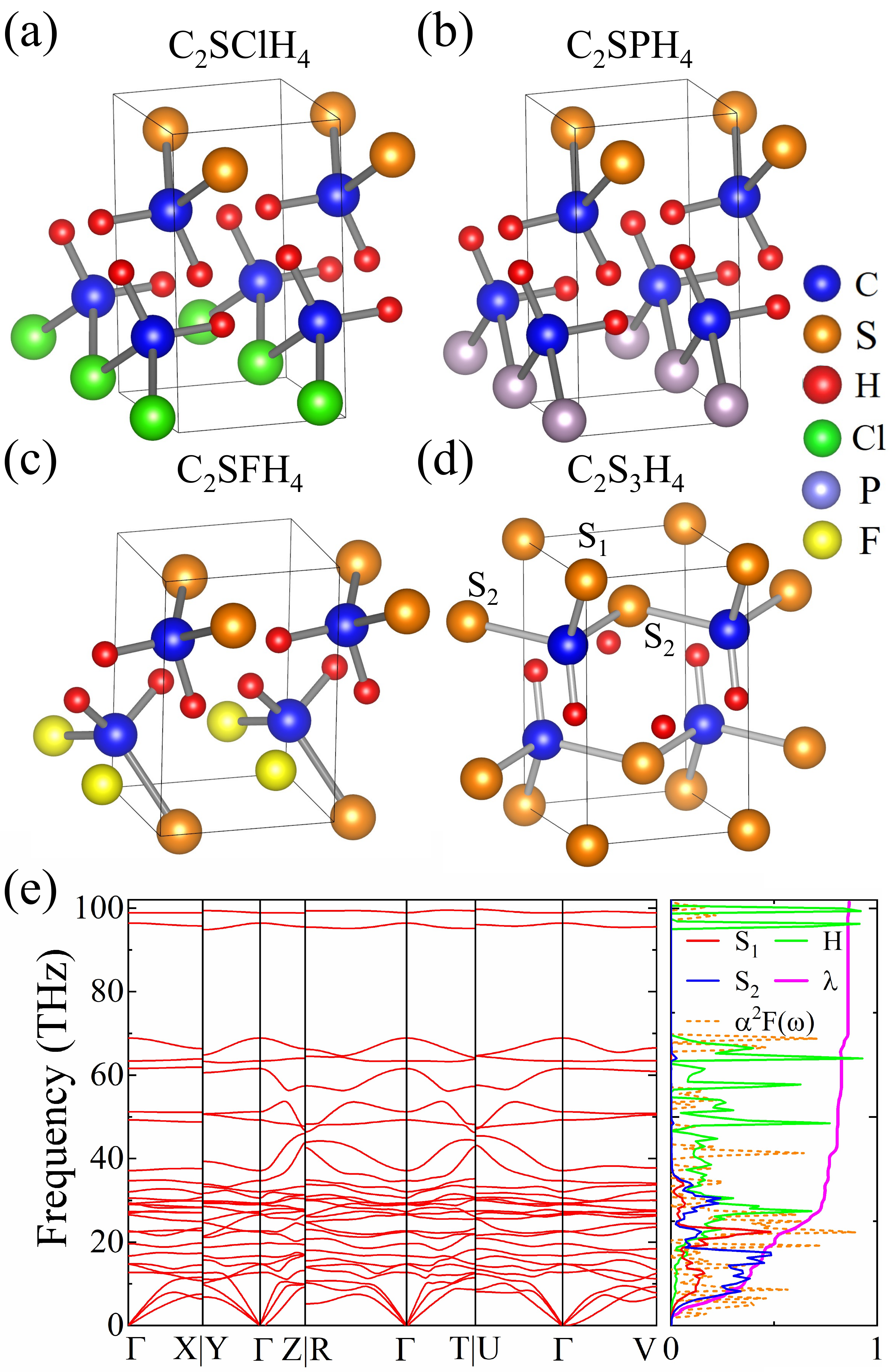}\\
	\caption{The crystal structures of (a) C$_{2}$SClH$_{4}$: replacing a S atom with a Cl atom, (b) C$_{2}$SPH$_{4}$: replacing a S atom with a P atom, (c) C$_{2}$SFH$_{4}$: replacing a S atom with a F atom, and (d) C$_{2}$S$_{3}$H$_{4}$: adding a S atom to the vertex of C$_{2}$S$_{2}$H$_{4}$. (e) The phonon spectra, projected phonon density of states, $\alpha^2F(\omega)$, and EPC $\lambda(\omega)$ of C$_{2}$S$_{3}$H$_{4}$ at 300 GPa.}\label{fig6}
\end{figure}

The temperature dependence of the electronic specific heat of C$_{2}$S$_{2}$H$_{4}$ at 300 GPa is presented in Fig.~\ref{fig5}(a). It is seen that the specific heat C(T) has distinct behaviors. When T$<$50 K, C(T) is almost zero. At temperature region of 50 K$<$T$<$300 K, C(T)$\sim$T$^3$ (left upper inset), and at T$>$300K, C(T)$\sim$T. The result of C(T) suggests that below 300 K, the normal state shows a non-Fermi liquid behavior~\cite{Schofield1999}.

To further verify this observation, we have also studied the temperature dependence of Lorenz number defined by $L = \kappa_e/(\sigma T)$, as given in Fig.~\ref{fig5}(b), as well as the electrical $\sigma$ and thermal conductivity $\kappa_e$ over relaxation time $\tau$. It can be observed that at temperatures lower than 300 K, $L$ is not a constant, showing dramatical violation of Wiedemann-Franz law~\cite{Franz1853}, while at higher temperature it shows almost a constant, exhibiting a Fermi liquid behavior.
Therefore, C$_{2}$S$_{2}$H$_{4}$ at 300 GPa shows a non-Fermi liquid behavior at temperatures below 300 K, implying that the interactions between electrons in the normal state play essential roles at low temperature. 

Based on C$_{2}$S$_{2}$H$_{4}$, we have designed four new structures as shown in Fig.~\ref{fig6}: (a) electron doping by replacing a S atom with a Cl atom; (b) hole doping by replacing a S atom with a P atom; (c) replacing a S atom with a light element F atom; (d) adding a S atom to the vertex of the primitive cell of C$_2$S$_2$H$_4$, where the added S atom is labeled as S$_1$ and the original S atoms are labeled as S$_2$. These four new structures are all demonstrated to be dynamically stable and exhibit superconductivity at 300 GPa (See details at Supplementary Information). TABLE ~\ref{T-2} lists the Tc and EPC $\lambda$ of C$_{2}$S$_{2}$H$_{4}$ and the other four new superconductors. It is noted that the $\lambda$ of C$_{2}$S$_{3}$H$_{4}$ is much higher than that of C$_{2}$S$_{2}$H$_{4}$, and Tc of the former is about three times larger than that of the latter. To uncover possible mechanism to enhance Tc, we compared the crystal structure, PhDOS, and $\alpha^2F(\omega)$ of C$_{2}$S$_{2}$H$_{4}$ and C$_{2}$S$_{3}$H$_{4}$. For C$_2$S$_2$H$_4$ in Fig.~\ref{fig1}, one C atom connects two S and two H atoms. Whereas, one C atom in C$_{2}$S$_{3}$H$_{4}$ connects three S atoms and one H atoms with another H atom originally connected to C atom isolated. The PhDOS in Fig.~\ref{fig6}(e) shows that the isolated H atom contributes a new vibration peak within 25 and 30 THz, and the added S atom (S$_1$) vibrates around 20 THz. The lowering vibration modes of S and H atoms greatly enhance the $\lambda$ from 0.5504 of C$_{2}$S$_{2}$H$_{4}$ to 0.8646 of C$_{2}$S$_{3}$H$_{4}$, giving rise to a higher Tc of 47.439 K in C$_2$S$_3$H$_4$. Our findings from C$_2$S$_2$H$_4$ to C$_2$S$_3$H$_4$ may provide a guide to find C-S-H superconductors with higher or even room temperature Tc experimentally in near future.

To conclude, we have searched for stable C-S-H structures with a C:S ratio of 1:1, and found a stable structure C$_{2}$S$_{2}$H$_{4}$. The Tc at 300 GPa is calculated to be 16.475 K, which is mainly attributed to the low-frequency vibration of S atoms. With the applied pressure, an insulator-to-metal phase transition occurs, accompanied by the system into superconducting at 64 GPa. The electronic thermal conductivity and Lorentz number of C$_{2}$S$_{2}$H$_{4}$ show a non-Fermi liquid behavior below 300K.
In addition, we have designed four new stable structures on the basis of C$_{2}$S$_{2}$H$_{4}$: C$_{2}$SClH$_{4}$, C$_{2}$SPH$_{4}$, C$_{2}$SFH$_{4}$, C$_{2}$S$_{3}$H$_{4}$. Among them, C$_2$S$_3$H$_4$ exhibts the highest Tc of 47.439 K. By analysis, we find that the added S atom in C$_{2}$S$_{3}$H$_{4}$ changes the original bonding of C$_{2}$S$_{2}$H$_{4}$, and contributs to new low-frequency vibration modes along with the isolated H atom. The molar ratio of C : S lower than 1:1 in the C-S-H systems may be favorable to enhance Tc, which will be useful not only to figure out the structure of the C-S-H material with room temperature Tc in the experiment, but also to find new superconductors with higher or even room temperature Tc in future.

This work is supported in part by the National Key R\&D Program of China (Grant No. 2018YFA0305800), the Strategic Priority Research Program of the Chinese Academy of Sciences (Grant No. XDB28000000), the National Natural Science Foundation of China (Grant No.11834014), and Beijing Municipal Science and Technology Commission (Grant No. Z118100004218001). B.G. is supported in part by the National Natural Science Foundation of China (Grants No. 12074378 and No. Y81Z01A1A9 ), the Chinese Academy of Sciences (Grants No. YSBR-030, No. Y929013EA2 and No. E0EG4301X2), the University of Chinese Academy of Sciences (Grant No.110200M208), the Strategic Priority Research Program of Chinese Academy of Sciences (Grant No. XDB33000000), and the Beijing Natural Science Foundation（(Grant No. Z190011).



%

\end{document}